\documentclass{svjour3}
\smartqed
\usepackage{graphicx}
\usepackage{amsfonts}
\usepackage{amssymb}

\begin{document}

\title{Strong Coupling Constant from $\tau$ Decay within a
Dispersive Approach to Perturbative QCD}

\titlerunning{Strong coupling constant}        

\author{B.A. Magradze}

\institute{ B.A. Magradze \at
            Andrea  Razmadze Mathematical Institute of I. Javakhishvili Tbilisi State University,
             2, University st., 0186 Tbilisi, Georgia  \\
                            \email{magr@rmi.ge}}

\date{Received: date / Accepted: date}
\maketitle

\maketitle

\begin{abstract}
We present a new dispersive framework for the extraction of the
strong coupling constant $\alpha_s$ from $\tau$-lepton decays. A
new feature of our procedure is the use of the quark-hadron
duality on the limited region $s_{\rm d}<s<m_{\tau}^{2}$. The
duality point $s_{\rm d}$ and the $\overline{\rm MS}$ strong
coupling constant $\alpha_{s}(m_{\tau}^{2})$  are
self-consistently extracted  from the  $\tau$ data for the
non-strange vector spectral function. We use 2005 ALEPH and 1998
OPAL experimental data on the vector spectral function. We compare
the new framework with the contour improved perturbation theory up
to order $\alpha_s^{5}$. The new procedure yields systematically
lower values for $\alpha_s$. From the 2005 ALEPH data,  we obtain
$\alpha_{s}(m_{\tau}^{2})=0.308\pm 0.014_{\rm exp}\pm 0.005_{\rm
th}$ which corresponds to $\alpha_{s}(M_{\rm _{z}}^{2})=0.1170\pm
0.0018_{\rm exp}\pm 0.0007_{\rm th}\pm 0.0005_{\rm ev}$. The
extracted value for the duality point $s_{\rm d}$ is found
surprisingly stable against perturbation theory corrections
$s_{\rm d}= 1.71\pm 0.05_{\rm exp}\pm 0.00_{\rm th}\,\, {\rm
GeV^{2}}$. From the 1998 OPAL data, we obtain
$\alpha_{s}(m_{\tau}^{2})=0.290\pm 0.023_{\rm exp}$ and $s_{\rm
d}=1.68\pm 0.10_{\rm exp}\,\, {\rm GeV^{2}}$.

\end{abstract}

\section{Introduction}
\label{intro} As is well known, the inclusive hadronic decays  of
the $\tau$-lepton may be reliable studied within perturbative QCD
(see seminal work \cite{BNP} and the literature therein). A
general approach to analyzing the perturbative and
non-perturbative aspects of the $\tau$-system observables is the
renormalization group improved perturbation theory augmented with
the Wilson's Operator Product Expansion (OPE) \cite{shifman}. The
characteristic energy scale is relatively small, $ m_{\tau}\approx
1.8\,\rm GeV$ ($ m_{\tau}$ being the mass of the $\tau$-lepton).
Hence, the non-perturbative effects of QCD cannot be completely
ignored. The original study \cite{BNP} has shown   that they are
small and under control within the OPE. In the following years,
the inclusive hadronic quantities of the $\tau$ system have been
intensively exploited to precisely determine the strong coupling
constant $\alpha_s(m_{\tau}^{2})$.
This became feasible because the observables of the $\tau$ system
are  sensitive to the concrete value of $\alpha_s$ and the
accuracy of the experimental data for a variety of the observables
has been considerably improved (for recent review see
\cite{DHZ1}).

In past few years, the determination of the strong coupling
constant from non-strange hadronic $\tau$-data has received a
renewed interest. It was pointed out \cite{Malt} that there is not
good agreement between recent two highest precision low-energy
determinations of $\alpha_s$. These determinations come from the
finite energy sum rule (FESR)
 analysis of hadronic $\tau$ decay data \cite{DHZ} and from a
 lattice perturbation theory analysis of ultraviolet-sensitive
lattice observables \cite{Mason}.
\begin{eqnarray}
\alpha_{s}(M_{z}^{2})&=& 0.1212\pm 0.0011\quad(\tau\quad \rm{decay})\\
\label{latt} \alpha_{s}(M_{z}^{2})&=& 0.1170\pm
0.0012\quad(\rm{lattice}).
\end{eqnarray}
Moreover, different determinations of $\alpha_s$  from the same
$\tau$ data \cite{ALEPH,compilation,OPAL} are not fully consistent
within their mutual errors (see work \cite{CGP1} and the
literature therein). This discrepancy has stimulated a number of
new theoretical investigations on the application of the FESR in
$\tau$ decays (see works
\cite{Malt,DHZ,CGP1,CGP,Baikov,Nari,Nari1,Domin,Domin1,Domin2,Pich,Boito}.
The standard FESR technique based on the truncated OPE series has
been reconsidered. The small but still significant
non-perturbative effects have been included into analysis
\cite{Boito}. On the one hand, the impact  of the  higher order
terms of the OPE (neglected in the standard analyzes)   has been
estimated \cite{Malt,Domin,Domin1,Pich}. It was confirmed that
their influence on the extracted value of $\alpha_s$ is not small
in the separate vector and axial vector channels. To suppress
these contributions in the FESR  the so-called pinched weights
introduced \cite{Malt,Domin,Domin1,Pich}. On the second hand,
using the physically motivated model \cite{CGP1,CGP,Boito} the
impact of the non-perturbative corrections coming from the
possible duality violations (DVs) \cite{shifman1}  has been
estimated. In the separate vector and axial vector channels the
DVs was found to be appreciable (see recent work \cite{Boito} and
references therein). The pinched weights have also been  employed
to reduce the effects of DVs  \cite{Pich}. Possible
non-perturbative corrections to the FESR (direct instantons,
duality violation and tachyonic gluon mass) which cannot be
described within the OPE have been estimated in \cite{Nari}.

As is well known, in the time-like region the renormalization
group (RG) invariance cannot be used unambiguously.  For this
reason, two different methods are used to perform the RG
resummation within the FESR. These are  fixed order perturbation
theory (FOPT) and contour improved perturbation theory (CIPT)
\cite{pivo,DP}. These two approaches lead to differing results.
The values of $\alpha_s$ extracted from $\tau$ decays employing
CIPT have always been higher. A critical comparison of these two
approaches may be found in recent works \cite{J,BJ}. In \cite{BJ}
FOPT was approved as a better approximation to the true result. In
contrast, authors of \cite{Malt} and \cite{DHZ}  favored CIPT.

Note that the non-physical singularities of the perturbative
running coupling (the Landau pole problem) which occur at small
space-like momenta may, supposedly, deteriorate the extracted
values of the parameters \cite{pivo1}. In particular, CIPT suffers
from this shortcoming \cite{BJ}. To cope with this problem
dispersive or analytic approaches to perturbative QCD have been
developed
\cite{SS,MSS,Y,SS2,MSa,MSS1,S1,my2,cvetic,cvetic1,baku,baku1,baku2,prosp}.
In works \cite{MSS} and \cite{Y}, the $\tau$ lepton decay rate has
been analyzed within a simple and effective dispersive technique,
Analytic Perturbation Theory (APT) (for reviews see
\cite{SS2,baku2,prosp}). However, the minimal analytic QCD model
(the same APT) predicts, from the non-strange $\tau$ lepton decay
data, too large value for the strong coupling constant,
$\alpha_s(m_{\tau}^2)=0.403\pm 0.015$ \cite{Y}.  The advantages
and shortcomings of the three approaches to the $\tau$ decays
(FOPT, CIPT and APT) were thoroughly analyzed in \cite{MSa}. It
should be noted that  APT as well as its generalized versions
suggested more later \cite{MSS1,my2,cvetic,cvetic1,baku,baku1}
proved to be very useful from the phenomenological point of view.
  A remarkable feature  of these modified expansions is the better
convergence and improved stability property with respect to change
of the renormalization scheme. Nevertheless, one should keep in
mind that an analytic approach based only on perturbation theory
can not be defined unambiguously, in fact, there is not a unique
recipe for removing the Landau singularities from the running
coupling.

In our earlier work \cite{my} we have suggested a  dispersive
approach to analyze the   $\tau$ decay data. In contrast to CIPT,
the new approach is based on the improved approximations to the
Adler function which incorporate correct analyticity and RG
invariance properties of the exact function. Moreover, the
approximations correctly reproduce the required ultraviolet and
infrared properties of the exact Adler function. Another feature
of the new framework is the use of the quark-hadron duality in the
limited region $s_{\rm d}<s<m_{\tau}^{2}$.  The QCD scale
parameter $\Lambda_{\overline {\rm MS}}$ and the duality point
$s_{\rm d}$ may be  determined, self-consistently, from the
experimental data \cite{my}.

In the present article, we investigate the new framework more
thoroughly. We revise part of the results of work \cite{my}. We
present  a more accurate test of the convergence of the numerical
results in perturbation theory. The numerical value of the duality
point $s_{\rm d}$ is found to be remarkable stable with respect to
higher order QCD corrections.
More importantly, we study the stability of the results with
respect to small change of the experimental data. In Sect. 2 we
critically analyze the FOPT and CIPT approaches to the
$\tau$-decay. A dispersive modification of the CIPT suggested in
\cite{my} is discussed in more detail. In Sect.~3 we give
corrected numerical values for $\alpha_s$ and $s_{\rm d}$
extracted from the 2005 ALEPH data. We thoroughly investigate  the
stability of the results comparing the new and CIPT determinations
of $\alpha_{s}$ order by order in perturbation theory. In
addition, we analyze the  ALEPH non-strange data employing the
renormalization scheme invariant (RSI) framework suggested  in
\cite{KKP}. We also analyze  1998 OPAL \cite{OPAL} vector data
within the new dispersive framework. Conclusions are summarized in
Sect.~4.

\section{Theoretical Framework}

Let us briefly recall some basic facts about the QCD analysis of
the hadronic decays of the $\tau$-lepton through  the FESR
\cite{KPTav}. The non-strange vector component of the
$\tau$-hadronic width is determined as
\begin{equation}
\label{rate} R_{\tau,V}=6|V_{\rm ud}|^{2}S_{\rm
EW}\int_{0}^{m_{\tau}^{2}}w_{\tau}(s)v_{1}(s)d\,s,
\end{equation}
where
$$w_{\tau}(s)={1\over m_{\tau}^{2}}\left(1-{s\over
m_{\tau}^{2}}\right)^{2}\left (1+2{s\over m_{\tau}^{2}}\right),$$
$V_{\rm ud}$ is the flavor CKM matrix element, $S_{\rm EW}$
denotes a short-distance electroweak correction \footnote{In what
follows, we neglect the small additive electroweak correction
${\delta^{\prime}}_{EW}$.}  and $v_{1}(s)$ is the vector spectral
function defined through the  correlation function \footnote{ We
use the normalization of the spectral function with the naive
parton prediction $v_{1,\rm naive}=1/2$. }
\begin{equation}
\label{spectral} v_{1}(s)=2\pi {\rm Im}{\Pi_{{\bar u}d,V}}(-s).
\end{equation}
It is more convenient to define a renormalization scale invariant
quantity, the Adler function
\begin{equation}
D(Q^{2})=-4\pi^{2}Q^{2}{d\over d\,Q^{2}}{\Pi_{{\bar
u}d,V}}(Q^{2}),
\end{equation}
here, we have defined $s=q^{2}=-Q^{2}$.  In the exact theory, the
correlation function $\Pi_{{\bar u}d,V}(z)$ and the Adler function
are analytic functions except the cut running along negative
$z$-axis. This implies the FESR relation
\begin{equation}
\label{FESR}
 R_{\tau,V}=-{3\imath\over
4\pi}|V_{\rm ud}|^{2}S_{\rm EW}\oint_{-s_0-\imath
\epsilon}^{{-s_0+\imath \epsilon}}\left(1-{z\over
s_0}\right)\left(1+{z\over s_0}\right)^{3}D(z){d\,z\over \,z},
\end{equation}
here, the integration contour   is a circle of radius $s_0$
($s_0=m_{\tau}^{2}$).   In the case of massless quarks, the Adler
function has the perturbation theory expansion \cite{J}
\begin{equation}
\label{series1}
D(Q^{2})=\sum_{n=0}^{\infty}a_s(\mu^{2})^{n}\sum_{k=1}^{n+1}kc_{nk}L^{k-1}\qquad
{\rm where} \qquad L=\ln{Q^{2}\over \mu^{2} },
\end{equation}
$a_s(\mu^{2})=\alpha_s(\mu^{2})/\pi$ and $\alpha_s(\mu^{2})$
denotes the strong coupling constant normalized at the scale
$\mu$. It follows from the renormalization scale invariance of the
Adler function that only the coefficients $c_{n1}$ are
independent. All other coefficients are determined  in terms of
the $c_{n1}$ and $\beta$-function coefficients through the RG
equation \cite{J,BJ}. In practice the series (\ref{series1}) is
truncated at some finite order.

The approximations to the Adler function obtained by truncation of
the series (\ref{series1}) have correct analytical properties of
the exact function. In the case of FOPT, the series
(\ref{series1}) is inserted into contour integral (\ref{FESR}) and
integrated term-by-term. Afterwards, the normalization scale is
determined choosing $\mu=m_{\tau}$ \cite{J}. However, we could
start from the original formula (\ref{rate}) with perturbation
theory expansion for the spectral function. The expansion for the
spectral function is obtained by insertion series (\ref{series1})
into inversion formula (\ref{inverse}) (see below) and integrating
term-by-term. So, we could achieve the same result without using
the FESR relation (\ref{FESR}). Thus, within FOPT, formulas
(\ref{rate}) and (\ref{FESR}) are equivalent. However,  the
approximations to the Adler function employed within FOPT do not
describe correctly the asymptotic  behavior of the exact function
for $Q^{2}\rightarrow \infty$. In the standard analysis of the
$\tau$ data this fact is irrelevant. However, as we shall see
latter, this is not the case for the new framework accepted in
this paper.

Since the Adler function is the renormalization scale invariant
quantity, one may choose $\mu^{2}=Q^{2}$ in  series
(\ref{series1}). Thus, one obtains the RG improved expansion
\begin{equation}
\label{series2} D(Q^{2})|_{\rm RG}=1+d(Q^{2})|_{\rm
RG}=1+\sum_{n=1}^{\infty}d_{n}a^{n}_s(Q^{2})
\end{equation}
where $d_n=c_{n1}$, $a_s(Q^{2})=\alpha_s(Q^{2})/\pi$,
$\alpha_s(Q^{2})$ being the running coupling. The first two
coefficients in the expansion (\ref{series2}) are the
renormalization scheme invariant. The  known  coefficients  in the
$\overline{\rm MS}$ scheme for $n_f=3$ quark flavors take values
$d_1=1$, $d_{2}\simeq 1.6398$, $d_{3}\simeq 6.3710$ and
$d_{4}\simeq 49.0757$. The last coefficient was calculated
recently by  the authors of \cite{Baikov}   by using powerful
computational techniques. The approximations to the Adler function
constructed by truncation the series (\ref{series2}) have correct
ultraviolet asymptotical behavior ($d(Q^{2})\rightarrow 0$ as
$Q^{2}\rightarrow\infty$), however they violate the cut-plane
analyticity  of the exact Adler function due to the non-physical
Landau singularities of the perturbative running coupling.  One
may assume,  without loss of generality, that the running coupling
has only one Landau singularity located on the positive $Q^{2}$
axis \cite{my}. This is true for the asymptotic solutions and for
more accurate Lambert-W solutions to the RG equation
\cite{my3,ggk,my4}. Then we may derive (see \cite{my}) the
violated dispersion relation for the QCD correction to the Adler
function:
\begin{equation}
\label{VDR} d(Q^{2})|_{\rm RG}=d(Q^{2})|_{\rm APT}+d_{\rm
L}(Q^{2}),
\end{equation}
where the function $d(Q^{2})|_{\rm APT}$ satisfies the normal DR
\begin{equation}
\label{DR} d(Q^{2})|_{\rm
APT}={1\over\pi}\int_{0}^{\infty}{\rho_{\rm
eff}(\sigma)\over{\sigma+Q^{2}}}d\,\sigma,
\end{equation}
with   the effective spectral density  determined as
\begin{equation}
\label{specfun} \rho_{\rm eff}(\sigma)={\rm Im}\{d(-\sigma-\imath
\epsilon)|_{\rm RG}\}.
\end{equation}
It is to be noted here that the function
\begin{equation}
\label{Aimage}
 D(Q^{2})_{\rm APT}=1+d(Q^{2})|_{\rm APT}
\end{equation}
is the analytic image of the perturbative Adler function
determined in the sense of the Analytic Perturbation Theory (APT)
approach of Shirkov and Solovtsov \cite{SS,MSS}. The second term
in (\ref{VDR}) is  the contribution coming from the Landau
singularity. It is represented by the contour integral \cite{my}
\begin{equation}
\label{L-part} d_{\rm L}(Q^{2})=-{1\over
2\pi\imath}\oint_{C_{L}^{+}}{d(\zeta)|_{\rm
RGI}\over{\zeta-Q^{2}}}d\,\zeta,
\end{equation}
here, the integral is taken round the circle $\{\zeta:
\zeta=s_{\rm L}+s_{\rm L}\exp{(\imath\phi)},-\pi<\phi\leq \pi\}$
in the positive (anti-clockwise) direction, with $s_L$ being the
Landau singular point.

In the popular framework, referred to as contour improved
perturbation theory (CIPT) \cite{pivo,DP}, the (truncated)
expansion (\ref{series2}) is inserted into the FESR integral
(\ref{FESR}) and then integrated term by term. At this point, one
ignores the fact that with the approximation (\ref{series2})
formulas (\ref{rate}) and (\ref{FESR}) are not equivalent. Indeed,
the FESR relation (\ref{FESR}) can not be derived because of
violated analytical properties of the approximation
(\ref{series2}). This inadequacy repeatedly discussed in the
literature (see for example \cite{BJ} and \cite{MSa}).
Nevertheless, CIPT has been very successful  from the
phenomenological point of view. On the other hand, APT  is free
from this drawback. However, the analysis of the $\tau$ decay data
based on APT with massless quarks gave too large value for the
strong coupling constant \cite{Y}. Furthermore, in the infrared
region, the Adler function  can not be reproduced correctly within
APT, CIPT or FOPT. Thus, in APT the running coupling
$\alpha_{s}(Q^{2})$ has a finite limit as $Q^{2}\rightarrow 0$
\cite{SS}.  This leads to the apparent contradiction in the case
of the Adler function. In fact, the Adler function should vanish
at $Q^{2}=0$, as is manifested by Chiral Perturbation Theory
\cite{PPR}. In work \cite{MSS1}, APT has been modified by
considering the quark mass threshold effects for the light quarks.
In this way, correct descriptions of the Adler function and $\tau$
data was achieved. However, too large values for the effective
quark masses ($m_{\rm u}\sim m_{\rm d}\sim 330 \,\,{\rm MeV}$) was
predicted.

 As is well known, in the exact theory the  Adler function satisfies
 the dispersion relation (DR)
\begin{equation}
\label{Adler}
D(Q^{2})=Q^{2}\int_{0}^{\infty}{2v_{1}(s)ds\over(s+Q^{2})^{2}},
\end{equation}
the corresponding inversion formula reads
\begin{equation}
\label{inverse}
 v_{1}(s)={1\over 4\pi\imath}\oint_{-s-\imath
\epsilon}^{-s+\imath \epsilon}{D(z)\over z}d\,z,
\end{equation}
where the path of integration, connecting the points $-s\mp\imath
\epsilon$ on the complex $z$-plane,   avoids the  cut running
along the real negative z axis. The integral being traversed in a
positive (anticlockwise) sense. From the violated DR (\ref{VDR}),
we may also derive  the integral representation
\begin{equation}
\label{VDR1} D(Q^{2})|_{\rm
RG}=Q^{2}\int_{-s_L}^{\infty}{2v_{1}^{\rm
RG}(s)ds\over(s+Q^{2})^{2}},
\end{equation}
where the singular integral at the lower bound should be treated
in the sense of distribution theory \footnote{We have confirmed
formula (\ref{VDR1}) in the case of the one-loop order
$\beta$-function. }. It is to be noted that the spectral function
$v_{1}^{\rm RG}(s)$ may be again calculated via the inversion
formula (\ref{inverse}), but now the integration contour should
also avoid the non-physical cut running along the positive
interval $0<z<s_L$ (see Fig. 1).

 \begin{figure}
\centering
\includegraphics[scale=1]{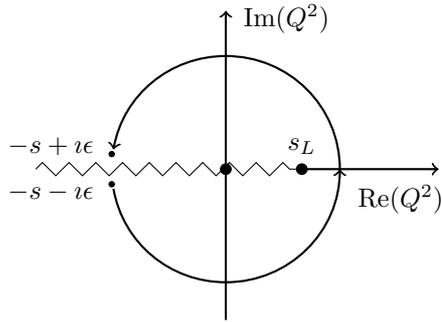}
\caption{The integration contour in the complex $Q^{2}$ plane used
in the inversion formula  (\ref{inverse}) in the case of the
approximation $D(Q^{2})|_{\rm RG}$ violating the DR.
 Branch points on the real axis are represented by the blobs and branch cuts
 by the zigzagging  lines. $s_{L}$ denotes the Landau singularity.}

\end{figure}

The dispersion relation (\ref{Adler}) may be used to construct the
approximations to the Adler function with correct analyticity
properties.  To approximate the hadronic spectral function, one
may use the global duality {\it ansatz} employed previously in
works \cite{BLR,PPR}
\begin{equation}
\label{ansatz1 } v_{1}(s)=\theta(s_{\rm d}-s) v_{1}^{\rm
np}(s)+\theta(s-s_{\rm d})v_{1}^{\rm pQCD}(s),
\end{equation}
where $v_{1}^{\rm pQCD}(s)$ is the perturbation theory
approximation to the spectral function, $v_{1}^{\rm np}(s)$
denotes the non-perturbative component of the spectral function
confined, presumably, in the low energy region, and $s_{\rm d}$ is
the onset of perturbative continuum, an infrared boundary in
Minkowski region above which we trust pQCD \footnote{It is assumed
that $0<s_{\rm d}<m_{\tau}^{2}$.}. One may also construct a
``semi-experimental''  spectral function
\begin{equation}
\label{GD} v_{1}^{\rm ``s.exp"}(s)= \theta(s_{\rm d}-s) v_{1}^{
 \rm exp}(s)+\theta(s-s_{\rm d})v_{1}^{\rm pQCD}(s),
\end{equation}
where $ v_{1}^{\rm exp}(s)$ denotes the genuine experimental part
of the total ``semi-experimental'' spectral function. It   was
measured with high precision by ALEPH \cite{ALEPH,compilation} and
OPAL \cite{OPAL} collaborations in the range
$0<\sqrt{s}<m_{\tau}=1.777\, {\rm GeV}$. Formula (\ref{GD})
extends the spectral function beyond the range accessible in  the
experiment. Formulas (\ref{ansatz1 }) and (\ref{GD}) provide
practical realizations of the concept of the quark-hadron duality
(see the original work \cite{BLR}). In \cite{PPR}, this {\it
ansatz} was used to determine the duality point $s_{\rm d}$ for a
given value of $\Lambda_{\overline{\rm MS}}$, the QCD scale
parameter in the $\overline{\rm MS}$ scheme. The perturbative
component $v_{1}^{\rm pQCD}(s)$ was constructed from the FOPT
series (\ref{series1}) choosing the normalization scale
$\mu^{2}=s_{\rm d}$. Such a framework may be considered as a
modification of  FOPT. In \cite{my}, we have used the same {\it
ansatz} for the spectral function. However, our strategy  was
somewhat different. Starting from the {\it ansatz} (\ref{GD}), we
have determined the parameters $\Lambda_{\overline{\rm MS}}$ and
$s_{\rm d}$ self-consistently from the $\tau$ data. In contrast to
\cite{PPR}, we have used the RG improved approximation $v_{1}^{\rm
RG}(s)$ to the spectral function. The function $v_{1}^{\rm RG}(s)$
is calculated by insertion of the (truncated) RG improved series
(\ref{series2}) into inversion formula (\ref{inverse}). For $s>0$,
one finds \cite{my}
\begin{equation}
\label{ptcomp} v_{1}^{\rm pQCD}(s)=v_{1}^{\rm RG}(s)=v_{1}^{\rm
APT}(s),
\end{equation}
where $v_{1}^{\rm APT}(s)$ is the spectral function determined in
the sense of the Shirkov-Solovtsov APT
\begin{equation}
\label{rs0} v_{1}^{\rm APT}(s)={1\over 2}(1+r(s)) \quad {\rm
where} \quad r(s)={1\over \pi}\int_{s}^{\infty}{\rho_{\rm
eff}(\sigma)\over\sigma}d\,\sigma.
\end{equation}

It follows from the duality relation (\ref{GD}) that one may
calculate in QCD perturbation theory the decay rate of the $\tau$
lepton into hadrons of invariant mass larger than $\sqrt{s_{\rm
d}}$
\begin{equation}
\label{rate1} R^{\rm QCD}_{\tau,V}|_{s>s_{\rm d}}=6|V_{\rm
ud}|^{2}S_{\rm EW}\int_{s_{\rm
d}}^{m_{\tau}^{2}}w_{\tau}(s)v_{1}^{\rm APT}(s)d\,s,=R^{\rm
exp}_{\tau,V}|_{s>s_{\rm d}}
\end{equation}
so that
\begin{equation}
\label{first} \Phi_{\tau}(s_{\rm d},{\Lambda}^{2})=\int_{s_{\rm
d}}^{m_{\tau}^{2}}w_{\tau}(s)v_{1}^{\rm APT}(s)d\,s =\int_{s_{\rm
d}}^{m_{\tau}^{2}}w_{\tau}(s)v_{1}^{\rm exp}(s)d\,s.
\end{equation}
Using  relation (\ref{rs0}), one may  express the left hand side
of (\ref{first}) in terms of the effective spectral density
\cite{my}
\begin{equation}
\label{Phi1}
\begin{array}{l}
 \displaystyle{\Phi_{\tau}(s_{\rm d},{\Lambda}^{2})=(1-{\hat
s}_{\rm
 d})^{3}(1+{\hat
s}_{\rm d}){(1+r(s_{\rm d}))\over
4}}\\
\qquad {}{}\displaystyle{-{1\over 4\pi}\int_{{\hat s}_{\rm
d}}^{1}y^{-1}(1-y)^{3}(1+y)\rho_{\rm eff}(m_{\tau}^{2}y) d\,y},
\end{array}\end{equation}
where ${\hat s}_{\rm d}=s_{\rm d}/m_{\tau}^{2}$.

Inserting the duality {\it ansatz} (\ref{GD}) into DR
(\ref{Adler}) one constructs the ``semi-experimental'' Adler
function
\begin{equation}
D(Q^{2})|_{\rm ``s.exp"}=D(Q^{2},s_{\rm d})|_{\rm
exp}+D(Q^{2},s_{\rm d})|_{\rm pQCD},
\end{equation}
where the experimental and QCD components of the Adler function
are determined by
\begin{equation}
\label{Adlerp} D(Q^{2},s_{\rm d})|_{\rm exp}=Q^{2}\int_{0}^{s_{\rm
d}}{2 v_{1}^{\rm exp}(s)d\,s\over(s+Q^{2})^{2}},\quad
D(Q^{2},s_{\rm d})|_{\rm pQCD}=Q^{2}\int_{s_{\rm
d}}^{\infty}{2v_{1}^{\rm pQCD}(s)d\,s\over(s+Q^{2})^{2}}.
\end{equation}
In general, the QCD component $v_{1}^{\rm pQCD}(s)$ may contain
the non-perturbative corrections coming from  the OPE as well as
the duality violating  terms \cite{Boito} not included into the
OPE. Intuitively, it seems to us that the non-perturbative
corrections are more essential in the region $0<s<s_{\rm d}$. In
what follows,  we will ignore these non-perturbative corrections
into QCD component of the spectral function and employ the
perturbative approximation (\ref{ptcomp}). The power suppressed
part of the ``semi-experimental" Adler function is defined as
\begin{equation}
\label{powsup} D(Q^{2})|_{\rm pw.s}=D(Q^{2})|_{\rm
``s.exp"}-D(Q^{2})|_{\rm RG},
\end{equation}
it  may be represented in the form \cite{my}
\begin{equation}
\label{powsup2} D(Q^{2})|_{\rm pw.s}=2\int_{0}^{s_{\rm
p}}K(Q^{2},s)( v_{1}^{\rm exp}(s)-v_{1}^{\rm APT}(s))d\,s-d_{\rm
L}(Q^{2}),
\end{equation}
where $K(Q^{2},s)=Q^{2}/(Q^{2}+s)^{2}$. Formula (\ref{powsup2})
enables us to derive the asymptotic expansion   at large $Q^{2}$
\begin{equation}
D(Q^{2})|_{\rm pw.s}\sim \sum_{n=1}^{\infty}\eta_{\rm
n}\left({\Lambda^{2}\over Q^{2}}\right)^{n}
\end{equation}
where $\Lambda\equiv\Lambda_{\overline{\rm MS}}$ is the QCD scale
parameter in the $\overline{\rm MS}$ scheme and the coefficients
$\eta_{\rm n}$ depend on the dimensionless ratios
$\Lambda^{2}/m_{\tau}^{2}$ and $s_{\rm d}/m_{\tau}^{2}$. In the
case of massless quarks, the gauge invariant operator of dimension
two can not be constructed. Hence, it follows that $\eta_1=0$.
This condition from the OPE leads to the equation relating  the
parameters $s_{\rm d}$ and $\Lambda$  with the experimental
spectral function \cite{my}
\begin{equation}
\label{second}
\Phi_{\rm as}(s_{\rm d},{\Lambda}^{2})={1\over m_{\tau}^{2}}\int_{0}^{s_{\rm d}}v_{1}^{\rm exp}(s)d\,s,\\
\end{equation}
where
\begin{equation}
\label{Phi2} \Phi_{\rm as}(s_{\rm d},{\Lambda}^{2})={{\hat s}_{\rm
d}\over 2}(1+r(s_{\rm d}))+{1\over 2{\pi m_{\tau}^{2}
}}\int_{0}^{s_{\rm d}}\rho_{\rm eff}(\sigma)d\,\sigma+{c_{\rm
L}\over2}{\Lambda^{2}\over m_{\tau}^{2}},
\end{equation}
with ${\hat s}_{\rm d}=s_{\rm d}/m_{\tau}^{2}$ and the coefficient
$c_{\rm L}$ is a positive number independent of $\Lambda$
\begin{equation}
\label{cL} c_{\rm L}=\Lambda^{-2}{1\over 2\pi\imath }\oint_{\rm
C_L^{+}}d(\zeta)|_{\rm RG}d\,\zeta={1\over 2\pi}{s_{\rm L}\over
{\Lambda}^{2}}\int_{-\pi}^{\pi}d(s_{\rm L}+s_{\rm L}e^{\imath
\phi})|_{_{\rm RG}}d\,\phi,
\end{equation}
here $s_{\rm L}$ being the Landau singularity of the running
coupling. It is proportional to $\Lambda^{2}$ \footnote{Analytic
expressions   for $s_{\rm L}$ in the $\overline{\rm MS}$ scheme up
to fourth order in perturbation theory may be found in
\cite{my4}.}. Numerical values of the coefficient $c_{\rm L}$
calculated in the $\overline {\rm MS}$ scheme  are listed in Table
\ref{tab:1}. In the calculations we have used  the approximations
to the Adler function of increasing order \footnote{We  use the
abbreviation ${\rm N}^{k}{\rm LO}$ to denote the order ${\cal
O}(\alpha_{s}^{k+1})$ approximation to the Adler function. }. All
approximations  have been constructed with the four-loop order
exact (numeric) running coupling. For the unknown ${\cal
O}(\alpha_{s}^{5})$ correction to the Adler function, we use the
geometric estimate $d_{5}=d_{4}(d_{4}/d_{3})=378$ \cite{DHZ}.

An important remark is in order here. The advantage of the
approximation $v_{1}^{\rm APT}(s)$ is that it correctly describes
asymptotic behavior of the exact function as $s\rightarrow
\infty$; in this limit $v_{1}^{\rm APT}(s)\rightarrow 1/2$. In
contrast,  the FOPT approximation $v_{1}^{\rm FOPT}(s)$ increases
with $s$ as a polynomial of $\ln{s}$. This shortcoming of FOPT is
irrelevant as far as the duality relation (\ref{first}) is
concerned. However, Eq.~(\ref{second}) depends on the ultraviolet
properties of the Adler function. This discussion suggests that a
more consistent framework should be constructed in the contour
improved scheme. In this work, we will refer the new framework  as
dispersive contour improved perturbation theory (DCIPT) \footnote{
In \cite{my}, we used the abbreviation $\rm APT^{+}$.}. Although
technically DCIPT resembles APT, there are significant differences
between the two frameworks. Thus, in DCIPT we do not mention
modifications of the QCD $\beta$-function and running coupling.

\begin{table}
\caption{Numerical values of the coefficient $c_{\rm L}$ in the
$\overline{\rm MS}$ scheme calculated with the four-loop order
exact numeric running coupling.} \label{tab:1}
\begin{tabular}{lccccc}\hline\noalign{\smallskip}
&\multicolumn{5}{c}{Approximations to the Adler
function}\\\cline{2-6}\noalign{\smallskip} & $\rm LO$& $\rm NLO$ &
$\rm N^{2}LO$ &$\rm N^{3}LO$&$\rm
N^{4}LO$\\\hline\noalign{\smallskip}
$c_L$& 0.301262& 0.453421& 0.555401& 0.651373& 0.721687\\
\hline\noalign{\smallskip}
\end{tabular}
\end{table}

\section{Numerical Results }
The parameters $s_{\rm d}$ and $\Lambda$ may be extracted from the
data by solving the system of equations
\begin{eqnarray}
\label{transc1} \Phi_{\tau}(s_{\rm
d},{\Lambda}^{2})&=&\int_{s_{\rm
d}}^{m_{\tau}^{2}}w_{\tau}(s)v_{1}^{\rm exp}(s)d\,s,\\
\label{transc2} \Phi_{\rm as}(s_{\rm d},{\Lambda}^{2})&=&{1\over
m_{\tau}^{2}}\int_{0}^{s_{\rm d}}v_{1}^{\rm exp}(s)d\,s,
\end{eqnarray}
where the functions $\Phi_{\tau}$ and $\Phi_{\rm as}$ are defined
in formulas (\ref{Phi1}) and (\ref{Phi2}). The right hand sides of
Eqs.~(\ref{transc1})-(\ref{transc2}) are determined in terms of
the empirical function $v_{1}^{\rm exp}(s)$. We have reconstructed
the experimental vector spectral function from the 2005 ALEPH
spectral data for the vector invariant   mass squared distribution
\cite{compilation}. This was done, with  the values
$|V_{ud}|=0.9746\pm 0.0006$ and $S_{\rm EW}=1.0198$ quoted in
\cite{ALEPH}. To interpolate the spectral function between the
fixed experimental values of the energy squared, we use cubic
splines. Evidently, the mean values of the parameters should be
determined from the mean value of $v^{\rm exp}_{1}(s)$. The error
analysis is based on the system of equations
(\ref{transc1})-(\ref{transc2}) \cite{my}. To determine the
experimental uncertainties on the extracted values of the
parameters, we use covariance matrices provided by ALEPH.
 Unfortunately, in the earlier work \cite{my}, we
used (inconsistently)    the $\rm N^{2}LO$ value $c_{\rm
L}=0.555401$ (see Table~\ref{tab:1}) in all other orders. In this
work, we present corrected results.

In general, the system (\ref{transc1})-(\ref{transc2}) has more
than one solution. For phenomenological reasons, we look for the
solution  in the limited region $280\,{\rm MeV}<\Lambda<420\, {\rm
MeV}$. In this region, the system has only one solution. In Table
\ref{tab:2}, we give the central values for the parameters
extracted from 2005 ALEPH data within the new (DCIPT) framework.
\begin{table}
\caption{Central  values for the  parameters in the $\overline{\rm
MS}$ scheme extracted from the 2005 ALEPH vector $\tau$ data
order-by-order within   DCIPT. These results correspond to the
four-loop order running coupling.} \label{tab:2}

\begin{tabular}{lccccc}
\hline\noalign{\smallskip}
Observable&\multicolumn{5}{c}{Approximation to the Adler
function}\\\cline{2-6}\noalign{\smallskip}&$\rm{LO}$ &$\rm{NLO}$&
$\rm{N^{2}LO}$& $\rm{N}^{3}\rm{LO}$&$\rm{N}^{4}\rm{LO}$
\\\noalign{\smallskip}\hline\noalign{\smallskip}
$s_{\rm d}$\,\,${\rm GeV}^{2}$&1.707&1.710&1.709&1.707&1.705\\
$\Lambda$\,\,${\rm GeV}$ & 0.486& 0.378& 0.348 & 0.332 & 0.323  \\
$\alpha_{s}(m_{\tau}^{2})$ &0.401 &0.337&0.321&0.313 &0.308
\\\noalign{\smallskip}\hline
\end{tabular}
\end{table}
Formally, we may write a series for the numerical value of the
coupling constant as follows
$$
\alpha_{s}(m_{\tau}^{2})|_{\rm{N}^{4}\rm{LO}}=\alpha_{s}(m_{\tau}^{2})|_{\rm{LO}}+\sum_{k=1}^{4}\Delta_{k},
$$
where
$\Delta_{k}=\alpha_{s}(m_{\tau}^{2})|_{\rm{N}^{k}\rm{LO}}-\alpha_{s}(m_{\tau}^{2})|_{\rm{N}^{k-1}\rm{LO}}$.
Using the numbers listed   in Table~\ref{tab:2}, we obtain the
series
\begin{equation}
\label{numseries1}
\alpha_{s}(m_{\tau}^{2})|_{\rm{N}^{4}\rm{LO}}^{\rm
DCIPT}=0.401-0.064-0.016-0.009-0.005.
\end{equation}
In \cite{my}, from the same data,   we have obtained the  CIPT
series
\begin{equation}
\label{numseries2}
\alpha_{s}(m_{\tau}^{2})|_{\rm{N}^{4}\rm{LO}}^{\rm
CIPT}=0.485-0.095-0.023-0.013-0.007.
\end{equation}
In Table~\ref{tab:7}, we give the changes (in percents) of the
leading term induced by the consecutive corrections in the DCIPT
and CIPT series. One sees that the DCIPT series (\ref{numseries1})
converges more rapidly.
\begin{table}
\caption{The changes of the leading term induced by the
consecutive corrections in  the series  (\ref{numseries1}) and
(\ref{numseries2}).} \label{tab:7}
\begin{tabular}{|l|c|c|c|c|}
\hline
 Perturbative orders&NLO&$\rm N^{2}LO$&$\rm N^{3}LO$&$\rm N^{4}LO$\\\hline
DCIPT&15.9\%&4.0\%&2.2\%&1.2\%\\\hline
CIPT&19.6\%&4.7\%&2.7\%&1.4\%\\
 \hline
\end{tabular}
\end{table}

In this paper,  we will estimate only so called  indicative
theoretical errors. These are  defined as a half of the last
retained term in the series \cite{KKP}. As pointed out in
\cite{KKP},  this definition of the error  is heuristic and
indicative. From the DCIPT series (\ref{numseries1}), we obtain
the estimates
\begin{eqnarray}
\alpha_{s}(m_{\tau}^{2})|_{\rm{NLO}}&=&0.337\pm 0.016_{\rm exp}\pm
0.032_{\rm th}\nonumber \\
\alpha_{s}(m_{\tau}^{2})|_{\rm{N}^{2}\rm{LO}}&=& 0.321\pm
0.016_{\rm exp}  \pm 0.008_{\rm th}\nonumber\\
\alpha_{s}(m_{\tau}^{2})|_{\rm{N}^{3}\rm{LO}}&=& 0.313\pm
 0.014_{\rm exp}\pm 0.004_{\rm th}\nonumber\\
 \alpha_{s}(m_{\tau}^{2})|_{\rm{N}^{4}\rm{LO}}&=&0.308\pm
0.014_{\rm exp}\pm 0.002_{\rm th}\pm (0.0045_{d_5}),
 \label{ierrors1}
\end{eqnarray}
here we have also included the experimental errors \footnote{We
use formulas for the error analysis  derived in \cite{my}.}. In
previous paper \cite{my},  we have found from the same data in the
case of CIPT
\begin{eqnarray}
\alpha_{s}(m_{\tau}^{2})|_{\rm{NLO}}&=&0.390 \pm 0.011_{\rm
exp}\pm
0.048_{\rm th}\nonumber \\
\alpha_{s}(m_{\tau}^{2})|_{\rm{N}^{2}\rm{LO}}&=& 0.367 \pm
0.009_{\rm exp}\pm 0.012_{\rm th}\nonumber\\
 \alpha_{s}(m_{\tau}^{2})|_{\rm{N}^{3}\rm{LO}}&=&0.354\pm
 0.008_{\rm exp}\pm
 0.007_{\rm th}\nonumber\\
 \label{ierrors2}
\alpha_{s}(m_{\tau}^{2})|_{\rm{N}^{4}\rm{LO}}&=&0.347\pm
0.008_{\rm exp}\pm 0.003_{\rm th}\pm (0.0065_{d_5}).
\end{eqnarray}
 The $\rm{N}^{4}\rm{LO}$ estimates in (\ref{ierrors1}) and
(\ref{ierrors2}) correspond to the  central value $d_5=378$.  The
additional theoretical error in the coupling constant induced from
the uncertainty in the fifth order unknown coefficient  $(d_5=378
\pm 378)$ takes the values $0.0045$ $(\approx 1.5\%)$ and $0.0065$
$(\approx 1.9\%)$ in the new  and standard extraction procedures
respectively. Comparing formulas (\ref{ierrors1}) and
(\ref{ierrors2}), one sees that within DCIPT the indicative
theoretical errors take smaller values. In contrast to this, the
experimental errors on the values of $\alpha_s$ increases by the
factor of 1.75 within the new procedure. It is remarkable that the
more reliable estimate of the theoretical error presented in
\cite{DHZ} within CIPT (at $\rm N^{4}LO$) is close to  our
estimate of the error presented in formula (\ref{ierrors2}).

Similarly, determining the  theoretical  and experimental errors
on the parameter $s_{\rm d}$, we find stable results
\begin{eqnarray}
s_{\rm d}|_{\rm NLO}&=&1.710 \pm 0.054_{\rm exp}\pm 0.002_{\rm th}\,\,{\rm GeV}^{2} \nonumber \\
s_{\rm d}|_{\rm{N}^{2}\rm{LO}}&=&1.709\pm 0.054_{\rm exp}\pm 0.001_{\rm th}\,\,{\rm GeV}^{2} \nonumber\\
s_{\rm d}|_{\rm{N}^{3}\rm{LO}}&=&1.707\pm 0.054_{\rm exp}\pm
0.001_{\rm th}\,\,{\rm GeV}^{2}
 \nonumber\\
 \label{ierrors3}
s_{\rm d}|_{\rm{N}^{4}\rm{LO}}&=&1.705\pm 0.054_{\rm exp}\pm
0.001_{\rm th}\,\,{\rm GeV}^{2}.
\end{eqnarray}
It is seen from (\ref{ierrors3}), that the estimate for the
duality point $s_{\rm d}$   decreases very slowly with increasing
of the order of perturbation theory. Practically, it is constant,
$s_{\rm d}\approx 1.71\pm 0.05$\, $\rm GeV^{2}$.

Usually, it is convenient to perform evolution of the $\alpha_ s$
results to the reference scale $M_z=91.187\,\rm GeV$. This is done
by using RG equation and appropriate matching conditions at the
heavy quark (charm and bottom) thresholds (see \cite{rodrigo1} and
literature therein). The three-loop level matching conditions in
the $\overline{\rm MS}$ scheme  were derived in \cite{chetyrk1}.
In this paper, we  follow the work \cite{rodrigo2}. We perform the
matching at the matching scale $m_{\rm th}=2\mu_{\rm h}$ where
$\mu_{\rm h}$ is a scale invariant $\overline{\rm MS}$ mass of the
heavy quark $\mu_{\rm h}= {\overline m}_{\rm h}(\mu_{\rm h})$. We
assume for the scale invariant $\overline{\rm MS}$ masses the
estimates of the Particle Data Group $\mu_{\rm
c}=1.29^{+0.05}_{-0.11}\,\rm GeV$ and $\mu_{\rm
b}=4.19^{+0.18}_{-0.06}\,\rm GeV$ \cite{PPRD}. In the evolution
procedure, we use the exact numeric four-loop running coupling. In
Table~\ref{tab:4}, we compare the estimates for
$\alpha_{s}(M_{z}^{2})$ obtained from the two (DCIPT and CIPT)
$\tau$-decay determinations of the coupling constant.
\begin{table}
\caption{Comparison of the DCIPT and CIPT  $\tau$ decay
determinations of the strong coupling constant at the scale
$M_{\rm Z}=91.187\,\,{\rm GeV}$. Two errors are given, the
experimental (first number) and the error from the evolution
procedure (second number). } \label{tab:4}

\begin{tabular}{lcc}
\hline\noalign{\smallskip}
 Approximation&$\alpha_{s}(M_{z}^{2})|_{\rm DCIPT}$&
$\alpha_{s}(M_{z}^{2})|_{\rm{CIPT}}$\\

\noalign{\smallskip}\hline\noalign{\smallskip}
 $\rm{N}^{2}\rm{LO}$&$0.1187\pm
0.0019\pm 0.0005$&$0.1238\pm
0.0009\pm 0.0005$\\
$\rm{N}^{3}\rm{LO}$&$0.1176\pm 0.0018\pm 0.0005$&$0.1224\pm
0.0009\pm 0.0005$\\
 $\rm{N}^{4}\rm{LO}$&$0.1170\pm
0.0018\pm 0.0005$&$0.1217\pm
0.0009\pm 0.0005$\\

\noalign{\smallskip}\hline
\end{tabular}
\end{table}

As stated above, we have used the $\overline{\rm MS}$ scheme
four-loop running coupling uniformly in all calculations, whereas
the order of approximation to the Adler function has been varied
consecutively. To perform a more accurate test, let us now employ
the same orders to approximate  the $\beta$ and Adler functions.
The coefficient $c_{\rm L}$ is accordingly recalculated. In Table
\ref{tab:5}, we present the results of the improved test.
Comparing the numbers in Tables \ref{tab:2} and \ref{tab:5}, we
see that the extracted values for the parameters, beyond LO, are
very close (the $\rm N^{2}LO$ and $\rm N^{3}LO$ results
practically coincide).

\begin{table}
\caption{Testing the stability of the results with regard to
higher order perturbation theory corrections. Here, the
approximations to the $\beta$ and Adler functions are chosen
consistently, at the same orders.} \label{tab:5}

\begin{tabular}{l|cccc}
\hline\noalign{\smallskip}

 Perturbative order & $c_{\rm L}$ &$s_{\rm d}\,\,{\rm GeV}^{2}$&
$\alpha_{s}(m_{\tau}^{2})|_{\rm{DCIPT}}$\\

\noalign{\smallskip}\hline\noalign{\smallskip}
 LO&0.444444&1.721&0.394&\\
NLO&0.336798&1.713&0.335&\\
$\rm N^{2}LO$&0.527261&1.709&0.321&\\
$\rm N^{3}LO$&0.651373&1.707&0.313&\\
\noalign{\smallskip}\hline
\end{tabular}
\end{table}

Let us now   employ the renormalization scheme invariant
extraction method (RSI) of \cite{KKP} to extract  the numerical
values of the coupling constant from the 2005 ALEPH  V+A spectral
data.  We shall also include into consideration  the recently
calculated   ${\cal O}(\alpha_{s}^{4})$ term in the series
expansion of the Adler function. The advantage of this technique
is that one starts from the physical quantity, the effective
charge defined by
\begin{equation}
a_{\tau}={\alpha_{\tau}\over \pi}=\delta^{(0)}_{\rm th}
\end{equation}
where $\delta^{(0)}$ is the perturbative correction to  the
$\tau$-decay rate. The running coupling $a_{\tau}$  defines the
internal scheme for the physical quantity. The numerical value for
the QCD scale in the internal scheme, $\Lambda_{\tau}$, is
extracted using the equation
$a_{\tau}(m_{\tau}^{2})=~\delta^{(0)}_{\rm exp}$. The
$\overline{\rm MS}$ scheme scale parameter is determined according
to the relation $\Lambda_{\overline{\rm MS}
}=\Lambda_{\tau}\exp\{-5.20232/(2\beta_{0})\}$, where
$\beta_{0}=9/2$. Formulas for calculation of the coefficients of
the  function $\beta_{\tau}$ (the $\beta$-function in the internal
scheme)  may be found in works \cite{DHZ1,KKP}. For the
experimental value of the perturbative part of the $\tau$ decay
rate in the non-strange channel, we use the updated value
$$
\delta^{(0)}_{\rm exp}|_{V+A}=0.2042\pm 0.0050_{\rm exp},
$$
evaluated  recently in \cite{BJ}.    For consistency,  we use the
same orders to approximate the $\beta$  and     Adler functions in
the ${\overline {\rm MS}}$-scheme. In Table (\ref{tab:6}), we
compare, the RSI and DCIPT determinations of the coupling constant
order-by-order in perturbation theory. The relevant channels which
have been used to extract the coupling are indicated by
subscripts. It is seen from the Table, that beyond $\rm NLO$ the
two determinations of the coupling constant are in good agreement.
\begin{table}
\caption{Comparison of the RSI and DCIPT determinations of the
$\overline{\rm MS}$ coupling constant from the $\tau$-decay data.
Experimental errors are  given only.}

\label{tab:6}
\begin{tabular}{|l|l|l|}\hline
Perturbative order& $\alpha_{s}(m_{\tau}^{2})|_{V+A}^{\rm RSI}$&$\alpha_{s}(m_{\tau}^{2})|^{\rm DCIPT}_{V}$\\
&&\\\hline

$\rm NLO$ &$0.278\pm 0.003$ &$0.335\pm 0.016$ \\
  $\rm N^{2}LO$ &$0.319\pm 0.004$ & $0.321\pm 0.016$\\
$\rm N^{3}LO$& $0.312\pm 0.004$ & $0.313\pm 0.014$\\\hline
\end{tabular}
\end{table}

As is known, mathematically, the extraction of QCD parameters from
experimental data via sum rules constitutes a so called ill posed
inverse problem (analytical continuation of an approximately known
function) \cite{Domin2}. Small changes in the input data may lead
to large changes in the output. In this regard, it is desirable to
check  the new framework. To do such a test, one may extract the
values of the parameters using the data from different
$\tau$-decay experiments. As a different  experimental data, let
us employ 1998 OPAL experimental  data on the non-strange
isovector vector spectral function which is publicly available
\footnote{I would like to thank S. Menke and S. Peris for making
the data available to me.}. The data are arranged in 100 bins with
bin size $0.032\,{\rm GeV^{2}}$, starting from $s=0.016\,{\rm
GeV^{2}}$. Note that, the OPAL data correspond to the branching
fractions available in 1998, as well as the then-current values of
$V_{\rm ud}$ and the electronic branching fraction $B_{\rm e}$.
These parameters have been updated since then.  The 2005 ALEPH
analysis is more recent and based on more statistics. However,  it
was pointed out in \cite{Boito} that the correlations due to
unfolding have been omitted in the original ALEPH analysis. Thus
the publicly available covariance matrices \cite{compilation}
should be corrected. Fortunately, the above mentioned obstacles
have little relevance to the problem under investigation. In fact,
our aim is to investigate the impact of the  specific formulation
of the quark-hadron duality (as given in (\ref{GD})) on the
extracted value of $\alpha_s$.

Inserting into system of equations (\ref{transc1})-(\ref{transc2})
the empirical vector spectral function reconstructed from the 1998
OPAL data \footnote{We use cubic splines to interpolate the
data.}, we solve the system numerically. We use the ${\rm
N^{2}LO}$ and ${\rm N^{3}LO}$ approximations to the Adler function
combined for consistency with the three- and four-loop order
$\overline{\rm MS}$ running couplings respectively. To determine
the experimental uncertainties on the extracted values of the
parameters we use covariance matrices provided by OPAL  (relevant
formulas were derived in the appendix to \cite{my}). For the
duality point, we obtain stable result
\begin{eqnarray}
\label{dp-opal} s_{\rm d}|_{\rm N^2LO}&=&(1.680\pm
0.100_{\exp})\,{\rm
GeV^{2}}\nonumber\\
s_{\rm d}|_{\rm N^3LO}&=&(1.679\pm 0.100_{\rm exp})\,{\rm
GeV^{2}},
\end{eqnarray}
the central value in (\ref{dp-opal}) decreases very slowly as the
order in perturbation theory increases. We see that the numerical
values for the duality point extracted from the ALEPH and OPAL
data are close (cf. (\ref{ierrors3}) and (\ref{dp-opal}).   For
the strong coupling, from the OPAL data, we find the values
\begin{eqnarray}
\label{alp-opal}
\alpha_{s}(m_{\tau}^{2})|_{\rm N^2LO}&=&0.296\pm 0.025_{\rm exp}\nonumber\\
\alpha_{s}(m_{\tau}^{2})|_{\rm N^3LO}&=&0.290\pm 0.023_{\rm exp},
\end{eqnarray}
the central values here   are somewhat smaller as compared to the
corresponding values extracted from the ALEPH data (cf. formulas
(\ref{ierrors1}) and (\ref{alp-opal})). However, the two
determinations of the coupling constant are consistent within
their mutual errors. It should be remarked that, in the case of
the OPAL data, we have obtained larger experimental uncertainties
on the numerical values of the parameters. For comparison, the
original OPAL analysis   of the same data, within CIPT, gave the
value \cite{OPAL}
\begin{equation}
\label{alp-opal1} \alpha_{s}(m_{\tau}^{2})|_{\rm N^{3}LO}=0.348\pm
0.009_{\rm exp}\pm 0.019_{\rm th},
\end{equation}
Comparing the numbers in formulas (\ref{alp-opal}) and
(\ref{alp-opal1}), one sees that the DCIPT determination of
$\alpha_s$ is significantly smaller, and the two determinations
are not consistent within their mutual errors.
 Performing evolution of the  $\alpha_{s}$  values
(\ref{alp-opal}) to the $Z^{0}$-mass scale, we obtain
\begin{eqnarray}
\alpha_{s}(M_{z}^{2})|_{\rm N^2LO}&=&0.1154\pm 0.0034_{\rm
exp}\pm 0.0005_{\rm ev}\nonumber\\
 \alpha_{s}(M_{z}^{2})|_{\rm N^3LO}&=&0.1145\pm
0.0033_{\rm exp}\pm 0.0005_{\rm ev}.
\end{eqnarray}

\section{Conclusion}

We have extracted  numerical values for the $\overline{\rm MS}$
scheme strong coupling constant $\alpha_s$ from the $\tau$-lepton
decay   data. The data provided by   2005 ALEPH and 1998 OPAL
experiments are employed. We examine in detail the dispersive
approach to the $\tau$-decay suggested in our earlier work
\cite{my}.   The errors observed in some numerical results of
\cite{my} have been corrected. Accordingly some of the conclusions
of \cite{my} are changed.

The new framework is based on the approximations to the Adler
function which have correct analytical properties. So that the
application of the FESR (\ref{FESR}) is mathematically justified.
Moreover, these approximations correctly reproduce the infrared
and ultraviolet behavior of the exact Adler function. In contrast,
in the standard approaches (FOPT, CIPT or APT)  some of these
properties of the Adler function are violated. The global
quark-hadron duality is used in the limited region of the energy
squared $s_{\rm d}<s<m_{\tau}^{2}$ ($E_{\rm d}=\sqrt{s_{\rm
d}}\approx 1.31\,\rm{GeV})$. In the region $0<s<s_{\rm d}$, the
hadronic spectral function  is reconstructed from the experimental
data. This enabled us to reduce the effects of duality violations
coming from the low energy region. In fact, one expects in this
region  sizeable  non-perturbative corrections to the Adler
function.

Technically, the new method is based on the system of equations
(\ref{transc1})-(\ref{transc2}). The first equation follows from
the concept of global quark-hadron duality employed on the limited
interval of the energy squared, $s_{\rm d}<s<m_{\tau}^{2}$. The
second equation is a consequence of the OPE which imposes the
restrictions on the ultraviolet behavior of the Adler function.
The parameters $\alpha_s$ and $s_{\rm d}$ are simultaneously
extracted from the data. We have examined numerical stability of
the extracted values of the parameters order-by-order in
perturbation theory.  The new framework (DCIPT) and the standard
(CIPT) are systematically compared. We have demonstrated that the
DCIPT determinations of the strong coupling constant are more
stable against perturbation theory corrections (see
Table~\ref{tab:7}). The central value of the coupling constant
definitely became smaller as compared to the CIPT result (cf.
Eqs.~(\ref{ierrors1}) and (\ref{ierrors2})). The changes in the
central values are not within the quoted experimental and
theoretical errors. Using the error estimated within DCIPT,
$\sigma=\sqrt{\sigma_{\rm exp}^{2}+\sigma_{\rm th}^{2}}\approx
0.0151$, we find that at $\rm N^{3}LO$ the central values  of
$\alpha_s(m_{\tau}^{2})$ in formulas (\ref{ierrors1}) and
(\ref{ierrors2})  differ from each other in about 2.7 standard
deviation. However, assuming the error estimated within CIPT ,
$\sigma\approx 0.0107$, one finds even large difference,
$3.8\,\sigma$ \footnote{Due to the large experimental error within
DCIPT, $\sigma(\rm DCIPT)/\sigma(\rm CIPT)\approx 1.4$.}. A
shortcoming of the new procedure is the increased experimental
error on the extracted values of $\alpha_s$. This is a direct
consequence of the reduction of the duality region.

Having included into analysis the fourth order coefficient $d_4$
and the geometric estimate  $d_5=378$, we have observed excellent
agreement between the lattice and $\tau$-decay determinations of
the strong coupling constant. At $\rm N^{4}LO$,  the central value
for  $\alpha_{s}$  ( see Table \ref{tab:4}) coincides with the
central value quoted in \cite{Mason} (see formula (\ref{latt})).
For this reason, we believe that DCIPT provides better
approximation as compared to CIPT.

For comparison purposes, we have extracted the strong coupling
constant from the 2005 ALEPH V+A data by using the RSI method of
work \cite{KKP}, extending the result of \cite{KKP}  up to $\rm
N^{3}LO$ (see Table \ref{tab:6}). Good agreement  between the RSI
and DCIPT determinations of $\alpha_s$ has been observed.

The duality point $s_{\rm d}$ is found to be surprisingly stable
with respect to higher order QCD corrections: $s_{\rm d}=1.71\pm
0.05_{\exp}\pm 0.00_{\rm th}\,\,\rm GeV^{2}$ (see Tables
\ref{tab:2} and \ref{tab:5}). In Table \ref{tab:5},  we have
performed a more accurate  test of stability of the numerical
results, choosing consistently the orders of the  approximations
to the $\beta$- and Adler functions.

To examine the stability of the numerical results with respect to
change in the input data, we have also analyzed the 1998 OPAL data
for the non-strange vector spectral function. The extracted values
for the parameters from the ALEPH and OPAL data are found to be
consistent.

 The procedure suggested here can  obviously be extended for
analyzing  the non-strange $\tau$-data from the axial-vector (A)
and vector plus axial-vector (V+A) channels.

\begin{acknowledgements}
 It is a great pleasure to thank M.A. Eliashvili,
V. Gogokhia, V. Kartvelishvili, A.A. Khelashvili, A.N.
Kvinikhidze, H. Leutwyler, Z.K. Silagadze for helpful discussions
and comments. The present work has been partially supported by the
Georgian National Science Foundation under grant No
GNSF/ST08/4-400.
\end{acknowledgements}

{99}

\end{document}